\documentclass[twocolumn,aps,prl,superscriptaddress,letterpaper]{revtex4}

\newcommand{\ee}{\ensuremath{e^+e^-}}
\newcommand{\be}{\ensuremath{^8\mathrm{Be}^*}}
\newcommand{\beGround}{\ensuremath{^8\mathrm{Be}}}

\newcommand{\pr}[1]{\left( #1 \right)}

\usepackage{graphicx} 
\usepackage{amsmath}
\usepackage{comment}

\begin{document}

\title{A Standard Model Explanation for the ``ATOMKI Anomaly'' }
\author{A. Aleksejevs}
\affiliation{Grenfell Campus, Memorial University of Newfoundland, Corner Brook, NL, Canada}
\author{S. Barkanova}
\affiliation{Grenfell Campus, Memorial University of Newfoundland, Corner Brook, NL, Canada}
\author{Yu.G. Kolomensky}
\affiliation{Department of Physics, University of California, Berkeley, CA, USA}
\affiliation{Nuclear Science Division, Lawrence Berkeley National Laboratory, Berkeley, CA, USA}
\author{B. Sheff}
\affiliation{Department of Physics, University of Michigan, Ann Arbor, MI, USA}
 \date{\today}

\begin{abstract}
\noindent
Using the $e^+e^-$ pair spectrometer at the 5 MV Van de Graaff accelerator at the Institute for Nuclear Research, Hungarian Academy of Sciences (ATOMKI), Krasznahorkay {\it et al.} have claimed a 6.8$\sigma$ excess at high $e^+e^-$ opening angles in the internal pair creation isoscalar transition $^8\mathrm{Be}(18.15)\to\,^8\mathrm{Be}\,e^+e^-$. 
A hypothetical gauge boson with the mass circa $17$ MeV, ``X17", has been proposed as an explanation for the excess. We show that the observed experimental structure can be reproduced within the Standard Model by adding the full set of second-order corrections and the interference terms to the Born-level decay amplitudes considered by Krasznahorkay {\it et al}. We implement a detailed model of the ATOMKI detector, and also show how experimental selection and acceptance bias exacerbate the apparent difference between the experimental data and the Born-level prediction. 
\end{abstract}

\maketitle  


\section{Introduction}

The Standard Model (SM) of particle physics has survived many challenges over the past fifty years. Occasionally, experimental observations disagree with the SM predictions; such deviations are often reconciled after additional theoretical or experimental scrutiny. 
One such deviation is claimed by an experiment at the Institute for Nuclear Research, Hungarian Academy of Sciences (ATOMKI). In this experiment, a 5~MV Van De Graaf-accelerator was used to produce an 18.15 MeV excited state of 
$^8\mathrm{Be}(18.15)$ (called \be~henceforth) with a subsequent internal pair conversion (IPC) to the ground state of \beGround: $^7\mathrm{Li}(p,e^+e^-)\beGround$ \cite{Announcement}. This channel is of particular interest, as the creation of \ee\ pairs provides a potential avenue for detection of low-mass boson candidates that are not present in the Standard Model~\cite{Theory}. 
A pair spectrometer was constructed to focus on detecting the IPC  process with a fine resolution in the opening angle of \ee\ pairs  $\theta_{+-}$~\cite{Detector}. 

Recent results from this experiment indicate an excess of events at 
large $\theta_{+-}$ over the leading-order theoretical predictions, consistent with a 
new boson with a mass of $16.7$~MeV \cite{Announcement}. This experimental evidence was found to be inconsistent with a dark photon hypothesis~\cite{Theory}. Feng {\it et al.} proposed a new protophobic, fifth force gauge boson, which has sparked a flurry of additional model-building work and interest in popular press. In this paper, we discuss our effort to critically analyze the idea that a non-SM, resonant process is needed to produce the observed excess in \ee\ production at high $\theta_{+-}$. In order to fully understand the interplay between the theoretical and experimental effects, we construct a Monte Carlo (MC) model of the detector used at ATOMKI. We use this model to demonstrate that an interference between the Born-level IPC amplitude and a subleading nonresonant component (an amplitude with a broad phase-space structure) could produce effects observed by ATOMKI \cite{SheffThesis}. Similar ideas have been explored by other authors~\cite{Kalman:2020meg,Zhang:2017zap}. 
The most basic subleading contribution to the IPC process arises from the second order electromagnetic corrections, {\em i.e.} contributions beyond the Born approximation: two-photon box diagrams, vertex corrections, etc. While the calculation of the higher-order corrections is technically challenging, one would naively expect such contributions to be suppressed by a factor of $Z\alpha$ compared to the Born term. However, the interference between the box and tree-level diagrams and the nontrivial structure of the box diagrams, including kinematic singularities and the presence of excited nuclear states in the box, can produce unexpected effects. In this paper, we report the full second-order calculation of the 
$^8\mathrm{Be}^*\to\,^8\mathrm{Be}\,e^+e^-$ process, and compare the results to the experimental distributions reported by ATOMKI. 

\section{NLO QED Model of \be\ Decay}

In the rest frame of \be, the doubly-differential decay
rate comprises from a phase-space component and a square of the matrix
element, which has a leading-order (LO) contribution $\left|M_{LO}\right|^{2}$
and an interference term between LO and next-to-the-leading (NLO)
matrix element, $2\Re\left[M_{LO}M_{NLO}^{*}\right]$:

\begin{equation}
 \frac{d^{2}\Gamma}{d\theta_{+}d\theta_{-}}=\left(\left|M_{LO}\right|^{2}+2\Re\left[M_{LO}M_{NLO}^{*}\right]\right)\Phi,\label{tha1}
 \end{equation}
 where details on phase space element $\Phi$ are given in supplemental part of the paper. 

\begin{figure*}
\begin{centering}
\includegraphics[scale=0.7]{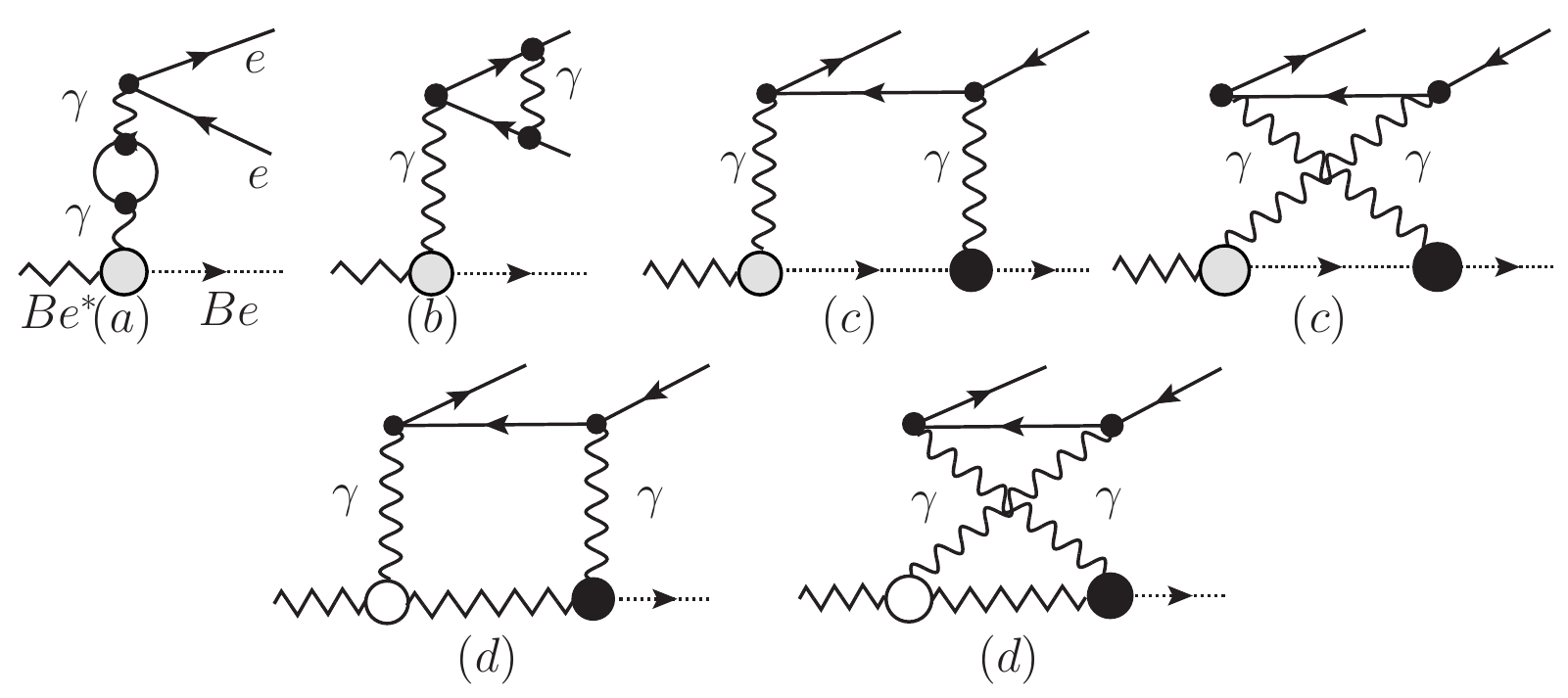}
\par\end{centering}
\caption{Higher order QED contributions in decay of \be. Grey bubble corresponds to spin $1\rightarrow 0$ transition, black bubble for spin $0 \rightarrow 0$ and white bubble for spin $1\rightarrow1$ transitions.}

\label{figQED1}
\end{figure*}

At the energy scale of the ATOMKI experiment, an effective operator
approach can be employed, with \be\ and \beGround\ as the fundamental
degrees of freedom. Specifically, for the transition\ $\be\rightarrow{}\beGround+\gamma$,
we can write:
\begin{align}
\mathfrak{L}^{(1)} & =\frac{e}{\Lambda_{\gamma}}\epsilon^{\mu\nu\alpha\beta}B_{\mu\nu}^{*}F_{\alpha\beta}B,\label{th1}
\end{align}
where $B_{\mu\nu}^{*}=\partial_{\mu}B_{\nu}^{*}-\partial_{\nu}B_{\mu}^{*}$
and $F_{\alpha\beta}=\partial_{\alpha}A_{\beta}-\partial_{\beta}A_{\alpha}$
are the usual field strength tensors for $^{8}\text{Be}^{*}$ and electromagnetic
fields, respectively. The constant $\Lambda_{\gamma}$ is related
to the transition nuclear matrix element and is used as a scaling
parameter. However, Eq.~(\ref{th1}) is only describing the tree-level
transition, which is insufficient to explain a small peak at $\sim16.7$
MeV of electron-positron invariant mass. Thus, we need to extend our
analysis from the Born (LO) matrix element of order $\alpha$ to the
next-to-leading-order (NLO) QED contributions, which requires one-loop
calculations. The four categories of Feynman graphs corresponding
to $\alpha^{2}$ transitions are shown on Fig.~\ref{figQED1}.
Group (a) in Fig.~\ref{figQED1} is the vacuum polarization contributions, which we treat by splitting the leptonic and hadronic contributions
and using effective masses of light quarks to calculate the hadronic contributions.
Group (b) in Fig.~\ref{figQED1}) is the electron current vertex corrections.
Groups (c) and (d) are represented by boxes with 
\beGround\ and \be\ 
in the loop, respectively. The diagrams (b)-(d)
are infrared-divergent. To treat infrared (IR) divergences in the
vertex correction graph (c), we use the soft-photon approximation
in bremsstrahlung diagrams and introduce a cut on energy of soft photons
given by the threshold conditions in the electron-positron pair production.
For the boxes, we take into account only the infrared-finite part. Since
ATOMKI does not distinguish between the electrons and positrons,
we sum over events in which either electron or positron are observed in the same detector element in the MC simulation. Due to the CP symmetry of the underlying QED lagrangian, 
this cancels out the IR-divergence in the box diagrams. 

Besides the QED transition $\be\rightarrow\beGround+\gamma$,
in the box diagrams, we also use
\begin{align}
 & \mathfrak{L}^{(2)}=-ieZ\left[B^{*\mu\nu}B_{\mu}^{*\dagger}A_{\nu}-B^{*\mu\nu\dagger}B_{\mu}^{*}A_{\nu}+B_{\mu}^{*}B_{\nu}^{*\dagger}F^{\mu\nu}\right]\label{th2}
\end{align}
to describe a coupling $\be\rightarrow\be+\gamma$,
and 
\begin{align}
 & \mathfrak{L}^{(3)}=-ieZ\left[B\partial_{\mu}B^{\dagger}-\partial_{\mu}BB^{\dagger}\right]A^{\mu}\label{th3}
\end{align}
to introduce coupling $\beGround\rightarrow\beGround+\gamma$. In both
lagrangians, $Z$ is equal to four. Calculations are done in several
stages. Based on the lagrangian densities in Eqs.~(\ref{th1}-\ref{th3})
we develop the model file, which we used in \texttt{FeynArts} \cite{feynarts} to generate one-loop topologies and produce matrix elements,
which we evaluate using Passarino-Veltman basis in the \texttt{FormCalc} \cite{frmcloop} package. Numerical calculations are carried out with the
help of \texttt{LoopTools} \cite{frmcloop} and confirmed by
\texttt{Collier} \cite{collier} packages.

\begin{figure}[h]
\begin{centering}
\includegraphics[width=0.95\columnwidth]{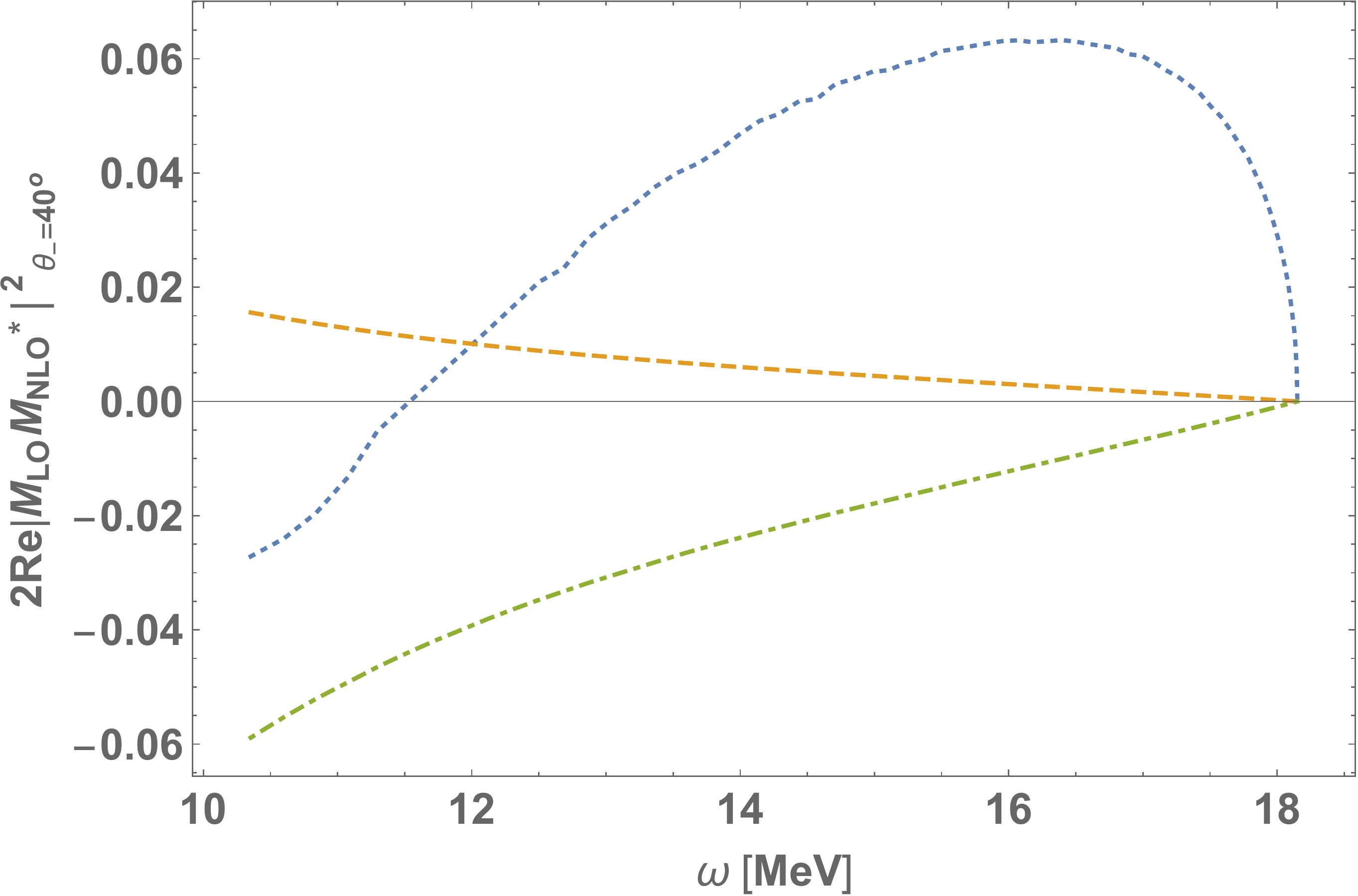} \includegraphics[width=0.95\columnwidth]{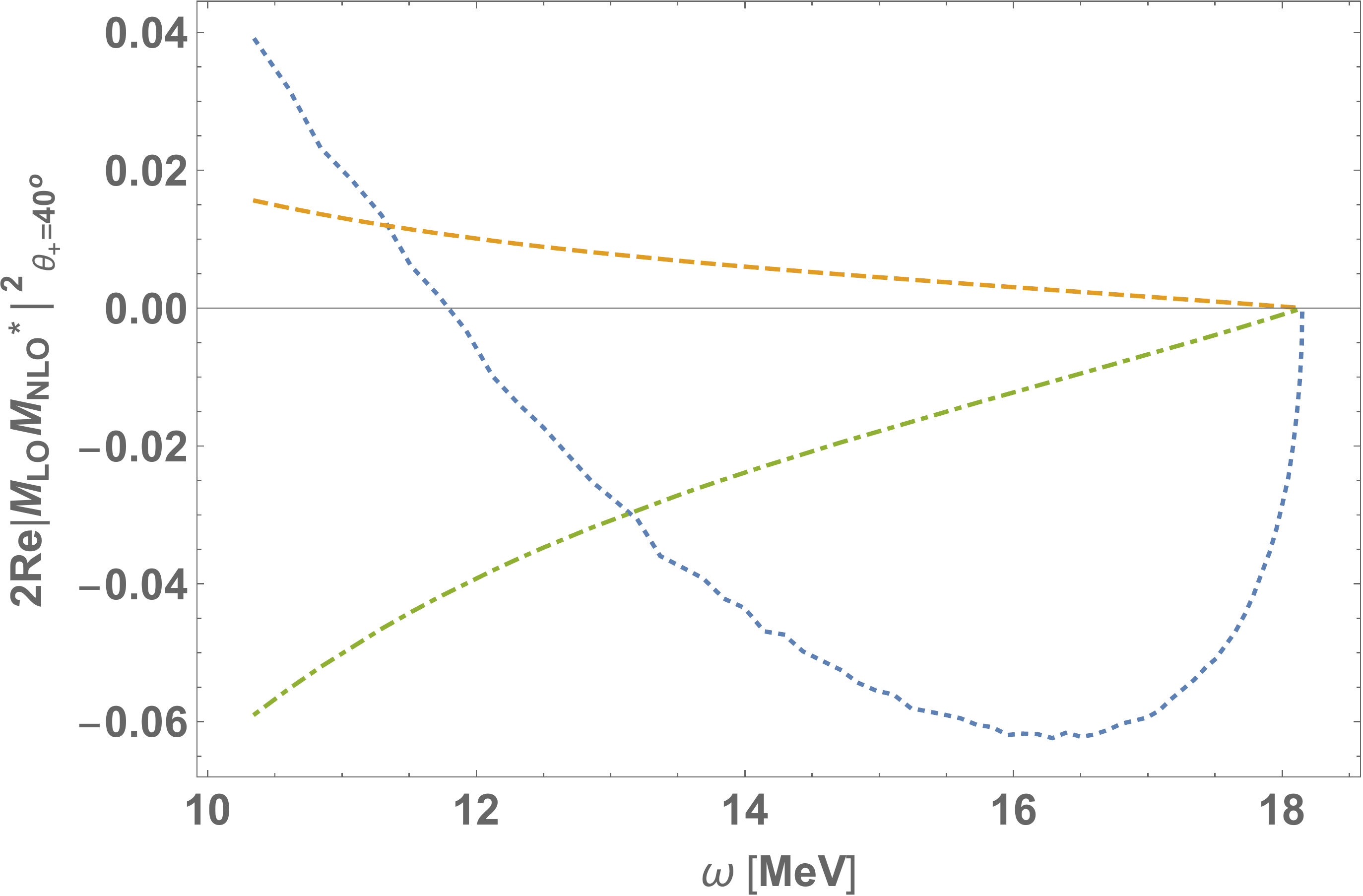}
\par\end{centering}
\caption{The interference term in Eq.~(\ref{tha1}) ($2\Re\left[M_{LO}M_{NLO}^{*}\right]$) as a function of $e^+e^-$ invariant mass $\omega$ for electron (top) and positron (bottom) angles fixed to $\theta_{-}(\theta_{+})=40^\circ$.
Dashed line (yellow) represents vacuum polarization graphs (a), dot-dashed
line (green) is the result of vertex correction graph (b) with soft-photon
bremsstrahlung treatment. Dotted graph (blue) is the IR-finite part
of the boxes (c)-(d).}

\label{figNLO}
\end{figure}
The results of the higher-order QED corrections for the cases when
one of the angles (either electron or positron) is fixed are given
in Fig.~\ref{figNLO}. Self-energies and vertex corrections are completely
symmetric with respect to electron-positron flip. On the contrary,
boxes show asymmetrical behavior and almost cancel out each other
when we combine the electron and positron events. The look-up tables
for the MC simulations are available upon request.

\section{Monte Carlo Simulation}

The momenta and angles of the final state particles \beGround, $e^+$, and $e^-$ are generated over the full three-body phase space with the probability proportional to the differential decay rate in Eq.~(\ref{tha1}). The kinematics are taken in the rest frame of \be, neglecting its boost due to the initial proton momentum. 
$e^+$ and $e^-$ are assumed to be indistinguishable in the detector, and we ignore the 4-vector of \beGround\ in the subsequent steps. 
The final momenta of \ee\ are then fed into the detector simulation and weighted accordingly. We then histogram over the opening angle and over the invariant mass of the \ee\ pair and compare against the experimental data. 

\begin{figure}[h]
   \centering
   \includegraphics[width=0.95\columnwidth]{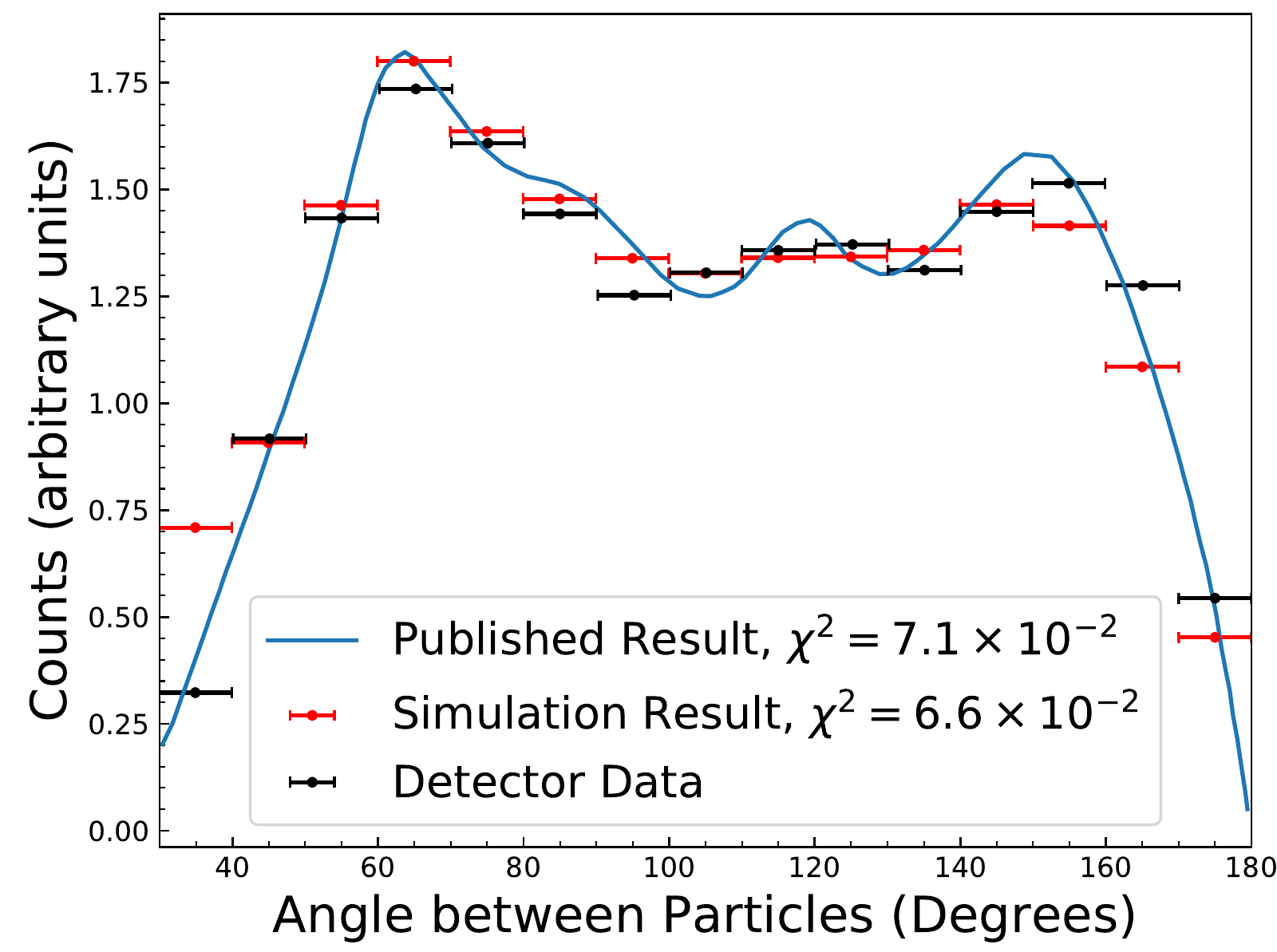} 
   \caption{\label{fig:responseCurve}Comparison of simulated and measured detector responses to uniformly distributed particle pairs, along with the published expected response. $\chi^2$ calculated with all standard errors assumed to be 1 for comparison of relative variance of new model and ATOMKI's model}
\end{figure}

The ATOMKI detector model was constructed based on description of \cite{Detector}. 
We assume each detector block to be a rectangle in $\theta\phi$ space with perfect efficiency. The center of each rectangle is placed at $\phi = 0, \pi/2, 5\pi/6, 7\pi/6, 3\pi/2$ \cite{Detector}. We tune the angular acceptance of each block by simulating a uniform angular distribution for each particle in \ee~pairs, and matching the angular acceptance of the \ee~pairs to the 
released detector response curve published  
\cite{Detector}, as shown in Fig.~\ref{fig:responseCurve}. The model represents the measured detector acceptance very well for angles larger than $40^\circ$, with comparable variance from data to the published expected detector response. The final block sizes are set to $\Delta\theta=0.59$ rad and $\delta\phi=0.58$ rad,  with the uncertainty of roughly $0.02$~rad in each dimension. These values are then applied to the MC simulation by applying a 0 weight to any \ee~pair that includes at least one particle not at an angle that crosses one of the detectors.

Based on \cite{Announcement}, an additional cut is applied to the detected particle pair, removing particularly asymmetric \ee~pairs. The cut was applied to only allow $|y| < 0.5$, where disparity $y = \left(E_{e^-} - E_{e^+}\right) / \left(E_{e^-} + E_{e^+}\right)$

We note that while $e^+$ and $e^-$ are indistinguishable in the ATOMKI detector, the response of the plastic scintillator to electrons and positrons is different~\cite{eePlastic}. It is not clear if this difference was incorporated in the original calibration of the ATOMKI apparatus~\cite{Detector}. 
We approximate the difference in response by adding an additional 1~MeV of visible energy to the positron signal before applying the disparity cut. This introduces a small asymmetry in the detector response. 

\section{Analysis}

\begin{figure}[h]
   \centering
      \includegraphics[width=0.95\columnwidth]{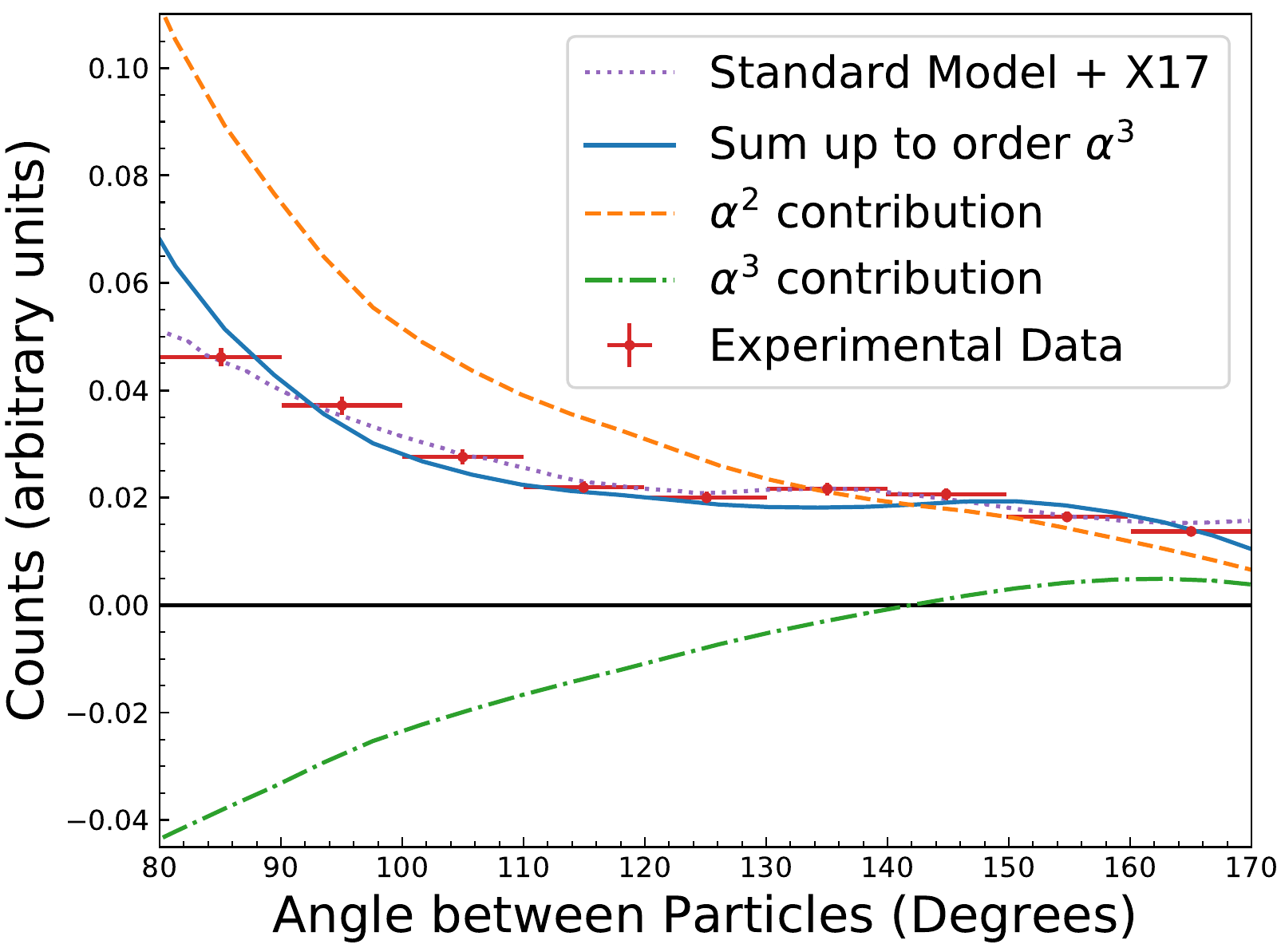}
      \caption{\label{fig:fit} Results of a Monte Carlo simulation using underlying particle distributions from the calculations above and applying detector acceptance and disparity limiting cuts. Shown are the (orange dashed) tree level calculation, (green dot-dashed) loop-level, (blue solid) their fitted sum, as compared to (purple dotted) the published result~\cite{Announcement} and (red points) the data collected in the same source.}
   \end{figure}

We fit the simulated distributions of \ee\ opening angles to the ATOMKI data using a binned $\chi^2$ fit by varying two parameters: the overall normalization of the Born contribution, and the coefficient of the interference term between the Born and the second-order diagrams. The sign and the magnitude of the interference term are allowed to float, allowing for a possible phase shift between the two terms. The best fit reproduces the ATOMKI spectrum with a $\chi^2=38$ for $7$ degrees of freedom. We note that the largest $\chi^2$ contribution comes from the lowest $\theta_{+-}$ bin, where our assumption of the uniform detector efficiency may be incorrect. The best-fit value for the interference coefficient is $\epsilon=57.72\pm 0.06$, indicating that the box diagram propagator may have contributions from more than one $J^P=1^+$ state ({\em e.g.} 18.15 MeV and 17.64 MeV states). 
This fit is shown in Fig.~\ref{fig:fit}.
   
   \vspace{-\baselineskip}
   \subsection{Mass Distribution}
   
   %
   %
   
   With a reasonable agreement between the SM fit and the experimental data in the X17 "signal" region at the large values of $\theta_{+-}$, we compute the \ee\ invariant mass distribution. First, for each of the two angular distributions used in the fit, a histogram of the \ee\ invariant mass is produced. 
   We use the X17 contribution reported in \cite{Announcement} as the detector mass resolution at $\bar{m}=16.7$~MeV, and scale that resolution with mass according to 
      \begin{equation}
      \text{g}_{m_0}(m) = \text{f}\pr{\frac{m - m_0}{m_0} \bar{m} + \bar{m}}
      \label{eq:massBlur}
   \end{equation}
where $m_0$ is the true $e^+e^-$ mass and $m$ is the measured mass. 
The theoretically predicted mass distribution is then smeared with this resolution function. 
   
\begin{figure}[h]
   \centering
      \includegraphics[width=0.95\columnwidth]{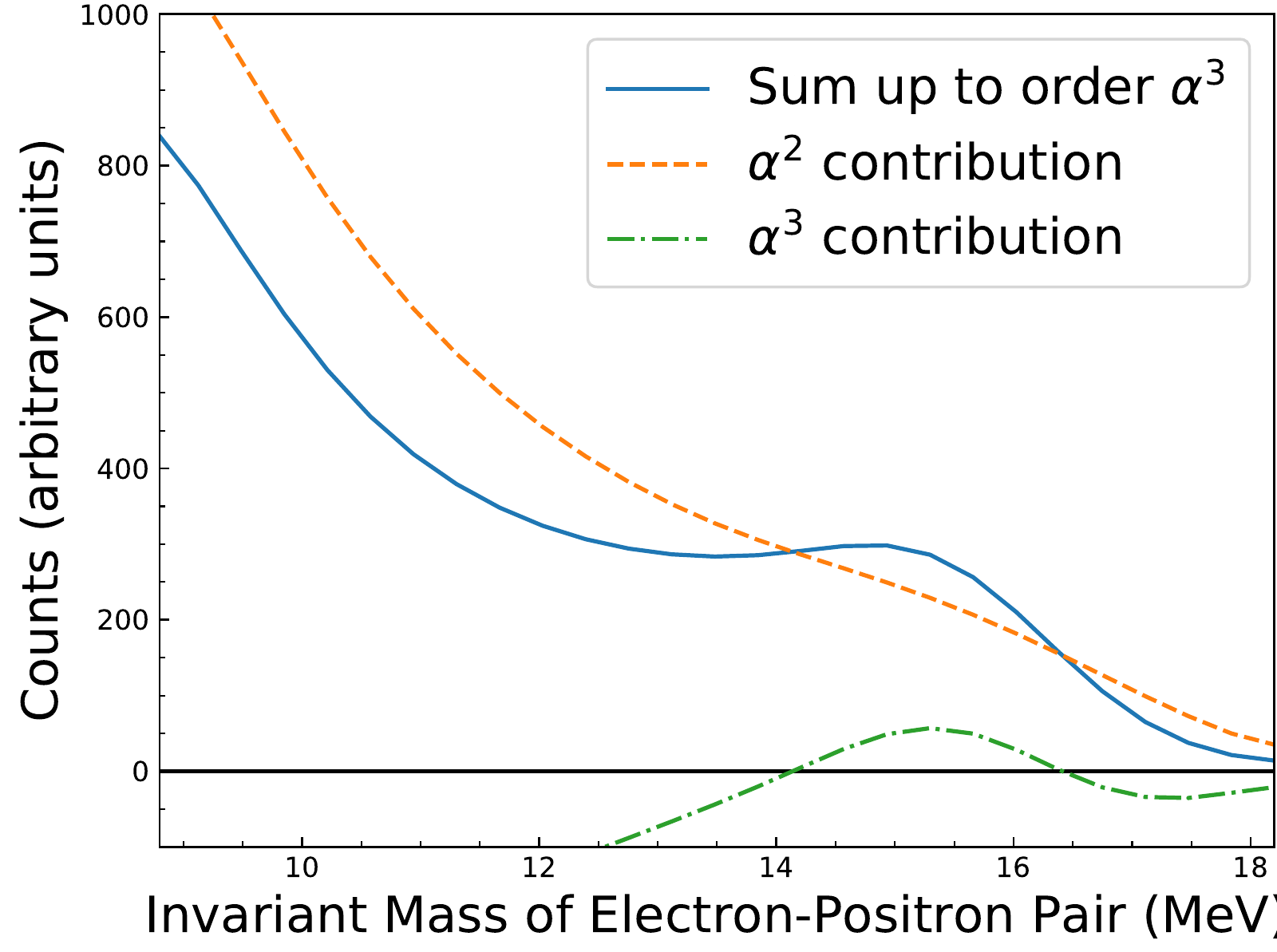}
      \caption{\label{fig:massfit} Distribution of electron positron pair creation over the invariant mass. Significant peak structure shown arising from interference between loop and tree level effects. Relative magnitudes of the components are taken from the fit in Fig.~\ref{fig:fit}.}
   \end{figure}

 The resulting mass distribution is shown in Fig.~\ref{fig:massfit}. We do not attempt to make a quantitative comparison with the ATOMKI mass distribution~\cite{Announcement}, since it depends critically on the detector energy resolution function, detection efficiency as a function of $e^-$ and $e^+$ energies, and details of the detector calibration. Such details are not provided in the experimental publications~\cite{Announcement,Detector}. However, we note a striking qualitative feature: a bump-like structure in the predicted invariant mass distribution that does not require a new boson. This shows how non-resonant SM processes can combine to produce a false bump in the detector.    
   
    \vspace{-\baselineskip}
  \subsection{Angular Distribution Bias}
\begin{figure}[h]
      \centering
      \includegraphics[width=0.95\columnwidth]{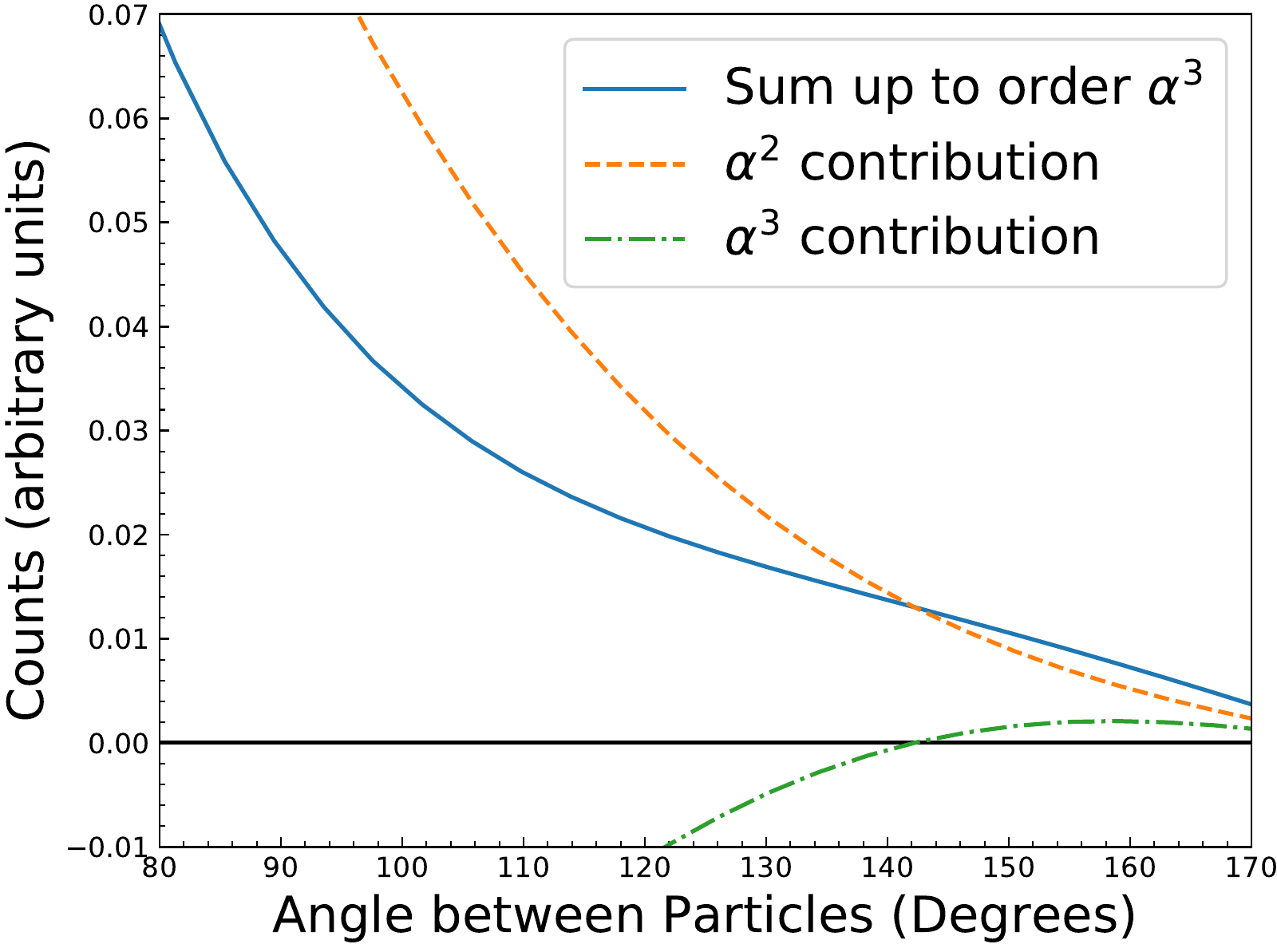} 
      \caption{\label{fig:noDetector} Angular separation distribution with detector set to full $4\pi$ acceptance. Energy disparity cuts still included. Shown are the (orange dashed) tree level calculation, (green dot-dashed) loop-level, and (blue solid) their sum. Relative magnitude of each component is taken from the fit found in Fig.~\ref{fig:fit}.}
  \end{figure}

\begin{figure}[h]
      \centering
      \includegraphics[width=0.95\columnwidth]{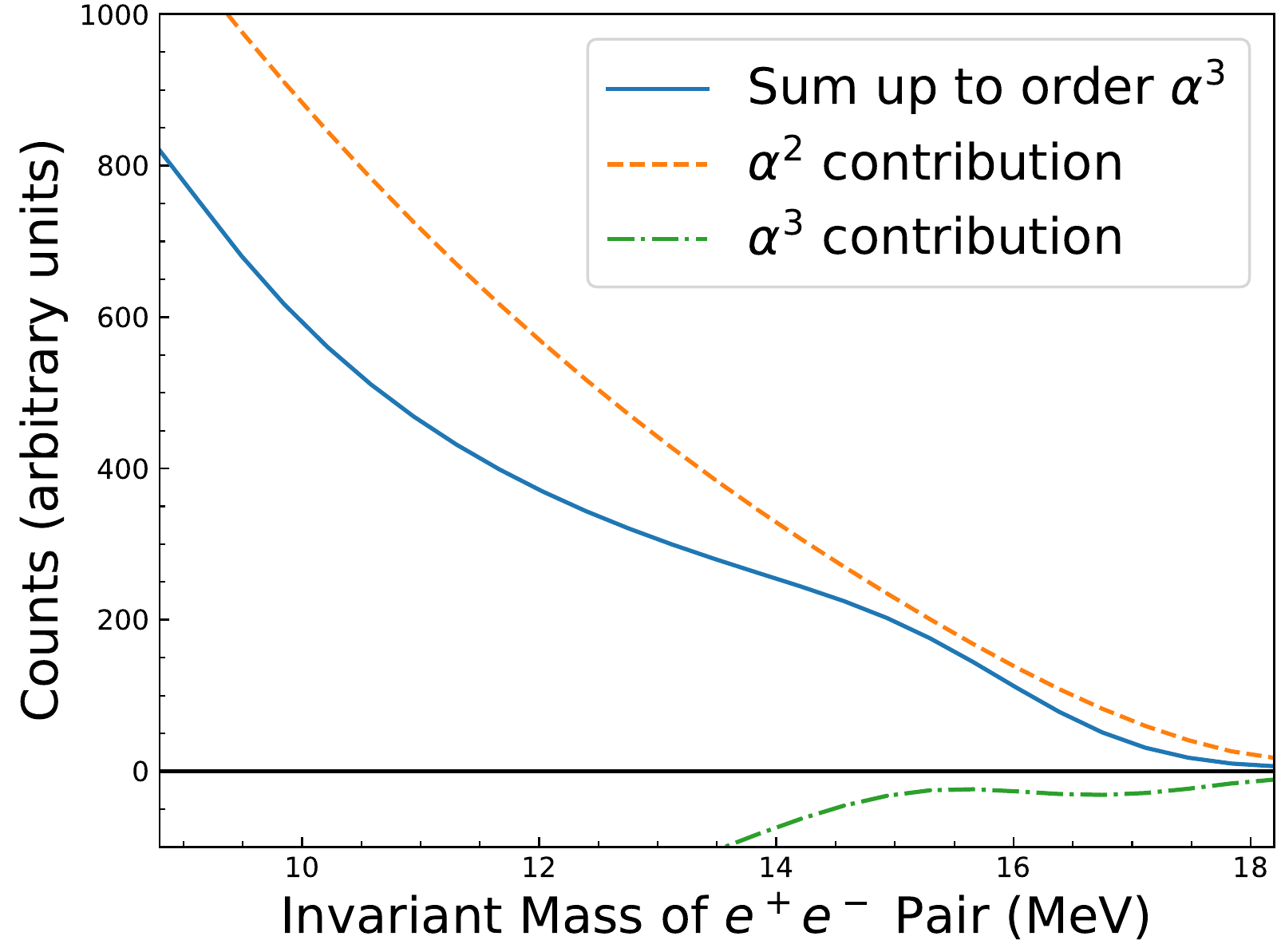} 
      \caption{\label{fig:noDetectorMass} Mass Distribution with the detector set to full $4\pi$ acceptance. Energy disparity cuts still included. Shown are the (orange dashed) tree level calculation, (green dot-dashed) loop-level, and (blue solid) their sum. Relative magnitude of each component is taken from the fit found in Fig.~\ref{fig:fit}}
  \end{figure}

   Of interest in this investigation are the origins of the bump-like structure in the angular and mass distributions in Fig.~\ref{fig:fit}-\ref{fig:massfit}. Much of this effect appears to arise from non-uniformities in the angular acceptance of the ATOMKI spectrometer. Running the simulation with complete $4\pi$ acceptance, and setting the interference coefficient to be the same as in the above fit, we arrive at the results shown in Fig.~\ref{fig:noDetector}-\ref{fig:noDetectorMass}. As shown, the bump structure is highly suppressed, implying that it is a result of an interplay between systematic biases in the detector and the effects of loop diagrams.
   
   \section{Conclusions}
   
   We show that the ATOMKI data \cite{Announcement} can be reasonably described by taking into account the second-order contributions to the IPC process, and accounting for detector and analysis bias. 
   As these are non-resonant, Standard Model processes, they do not require the addition of a new boson. Clearly, an element of caution must be taken before confidence can be placed in an entirely new particle having been discovered.
   
   An independent measurement of the IPC spectra in the \beGround\ and other isoscalar nuclear systems is clearly called for. Such measurements should aim to measure the IPC distributions with large acceptance, minimal detector bias, and ideally better invariant mass resolution, in order to be able to discriminate between a resonant and a non-resonant contributions. Tagging the initial nuclear state in order to eliminate initial-state interference effects~\cite{Kalman:2020meg}  would also be of value. 
   
   \section{Acknowledgments}
   
We would like to thank the members of UC Berkeley and Lawrence Berkeley National Lab (LBNL) Weak Interactions Group for their support in this work, and Gerald Miller and Xilin Zhou for stimulating discussions. This work was supported in part by the Natural Sciences and Engineering Research Council of Canada (NSERC), by the National Science Foundation (NSF), and by the US Department of Energy (DOE) Office of Science. 
B.~Sheff thanks UC Berkeley Institute for International Studies Merit Scholarship Program and UC Berkeley Physics department for their support. A.~Aleksejevs and S.~Barkanova would also like to thank UC Berkeley and LBNL for hospitality and support.

\bibliographystyle{plain}

\onecolumngrid
\newpage

\section{Appendix}

The phase space for doubly differential decay rate in  Eq.~(\ref{tha1}) can be written as follows:

 \begin{equation*}
 \Phi=\frac{1}{\left(2\pi\right)^{3}}\frac{J}{8m_{Be^{*}}}\frac{E_{-}^{2}}{E_{Be}}\frac{\sin\theta_{-}}{\sin\theta_{+}}
\end{equation*}
and
\begin{align}
 & J=\frac{\sin^{2}\left(\theta_{+}+\theta_{-}\right)}{2R\sin^{2}\theta_{-}}\left[\frac{m_{Be*}}{\left(\cos\left(\theta_{+}+\theta_{-}\right)-1\right)}\left(\frac{m_{Be^{*}}\sin\left(\theta_{+}+\theta_{-}\right)}{\tan\theta_{-}}-R\right)-m_{Be}^{2}\right],\nonumber \\
\nonumber \\
 & R=\left\{ m_{Be}^{2}+\frac{\sin\theta_{+}}{\sin\theta_{-}}\left(2m_{Be^{*}}^{2}+m_{Be}^{2}\frac{\sin\theta_{+}}{\sin\theta_{-}}\right)+\left(m_{Be^{*}}^{2}-m_{Be}^{2}\right)\frac{\sin\left(2\theta_{+}+\theta_{-}\right)}{\sin\theta_{-}}\right\} ^{1/2},\nonumber 
\end{align}
where subscripts $+/-$ correspond to the positron and the electron, respectively. 

\begin{figure}[hb]
\begin{centering}
\includegraphics[scale=0.2]{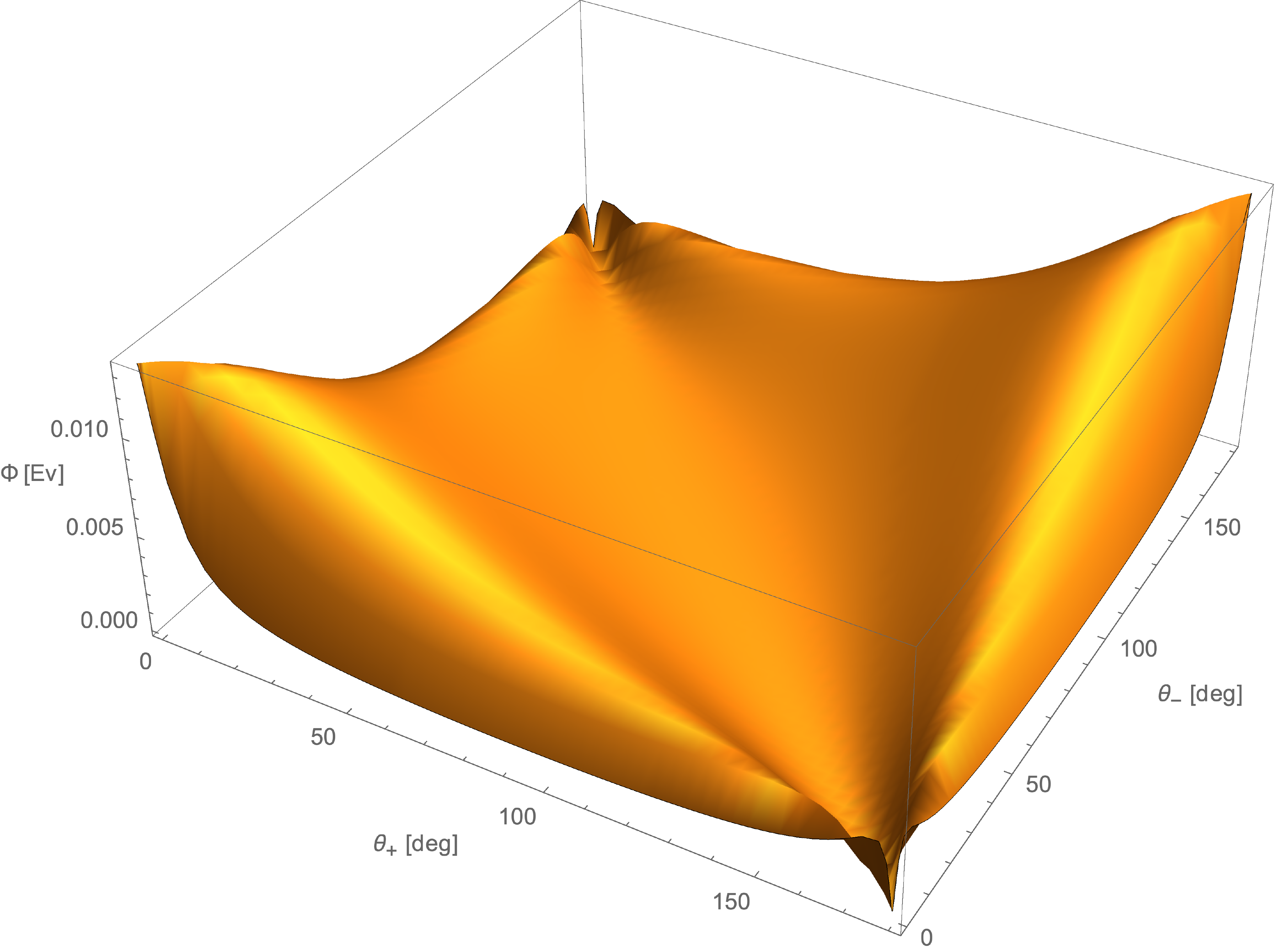} \includegraphics[scale=0.27]{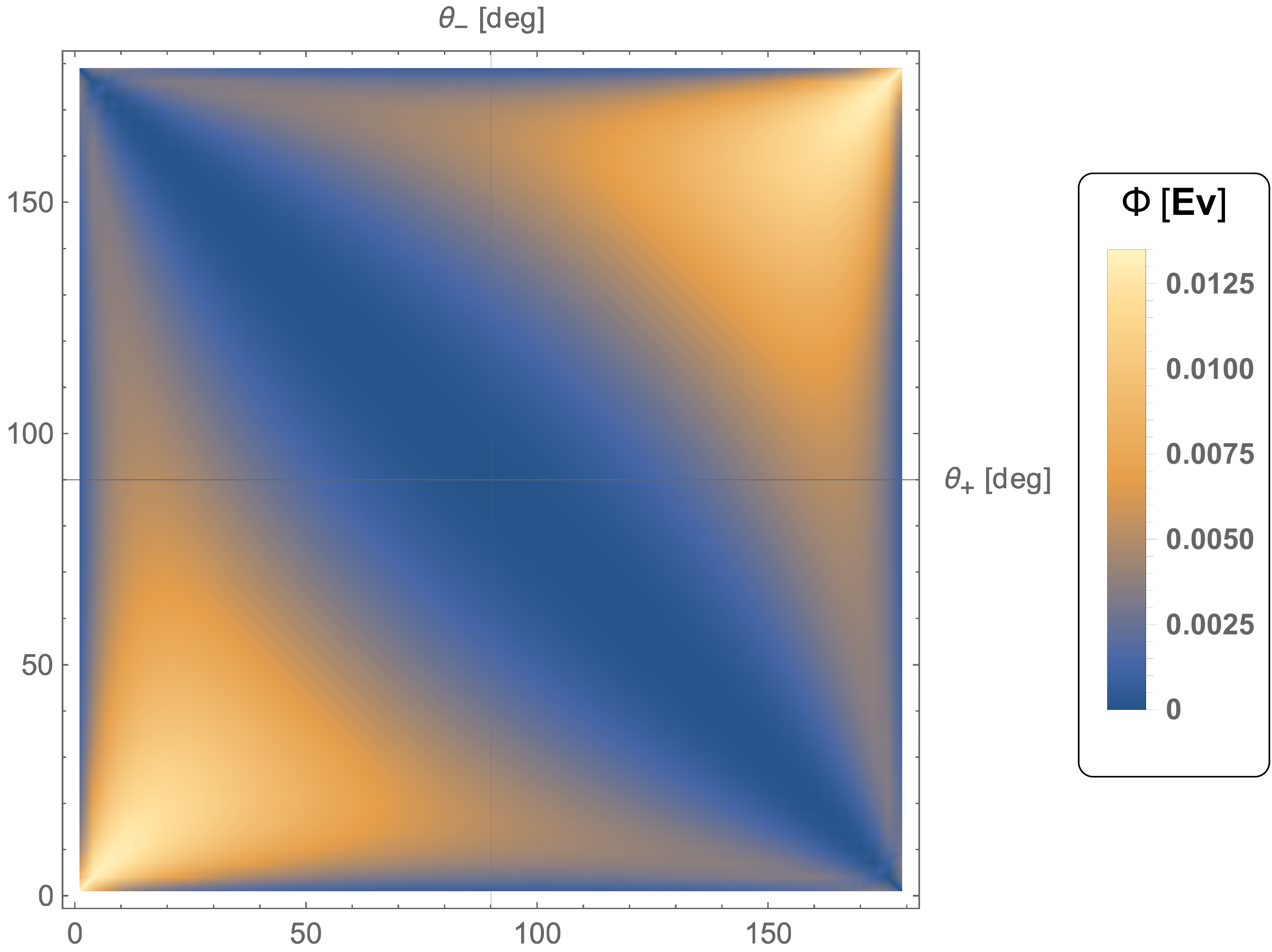}
\par\end{centering}
\caption{Phase space of three body decay for $\be\rightarrow\beGround e^{+}e^{-}$.}
\label{figPS1}
\end{figure}
  \begin{figure}[h]
      \centering
      \includegraphics[height=3in]{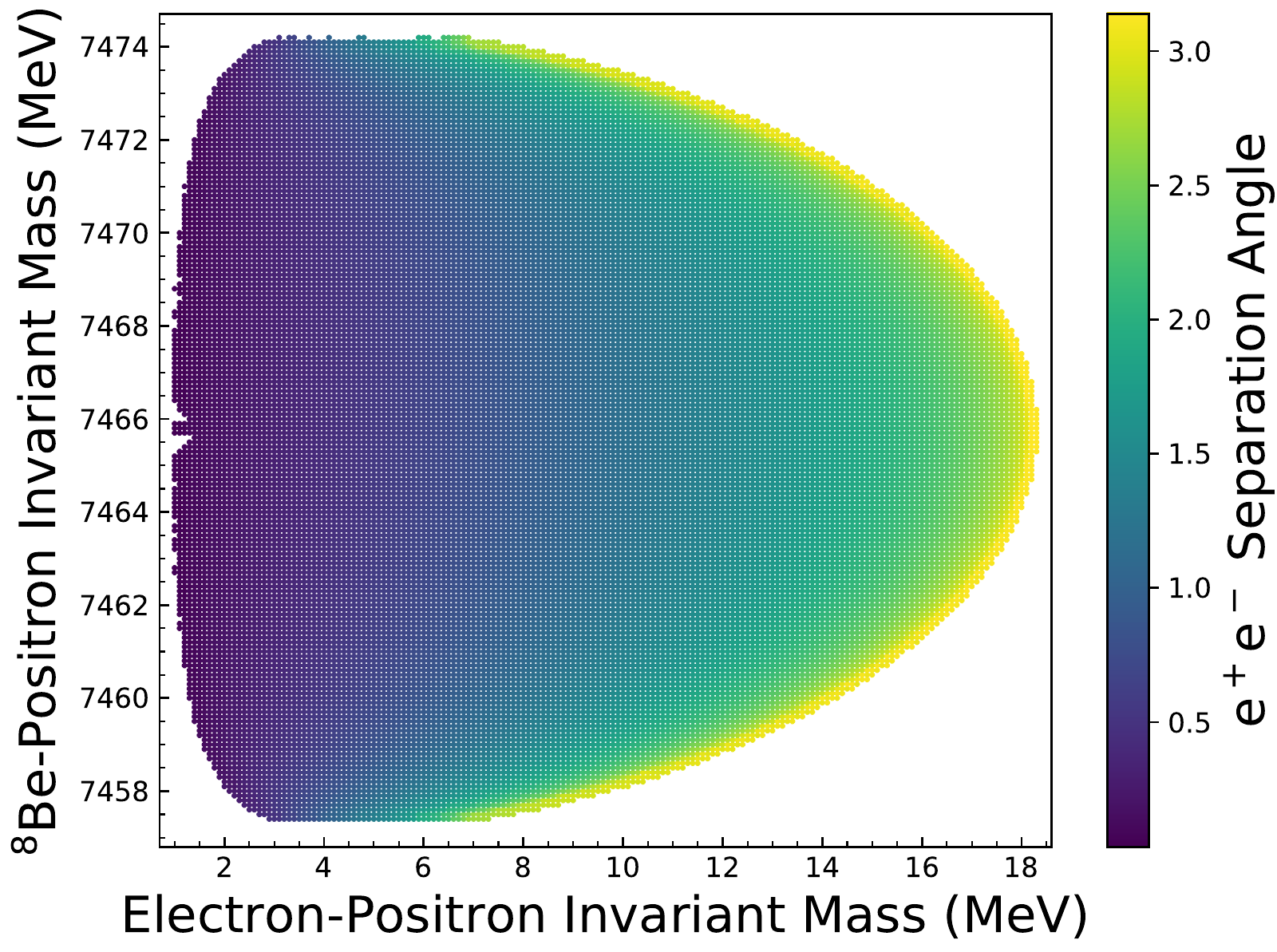}
      \caption{\label{fig:dalitzLinearangle} Separation angle between electron and positron as a function of invariant mass of the $\ee$ system and the $ e^+~\beGround$ system}
 \label{fig:angleDalitz}
  \end{figure}
\begin{figure}[h]
      \centering
      \includegraphics[width=0.48\textwidth]{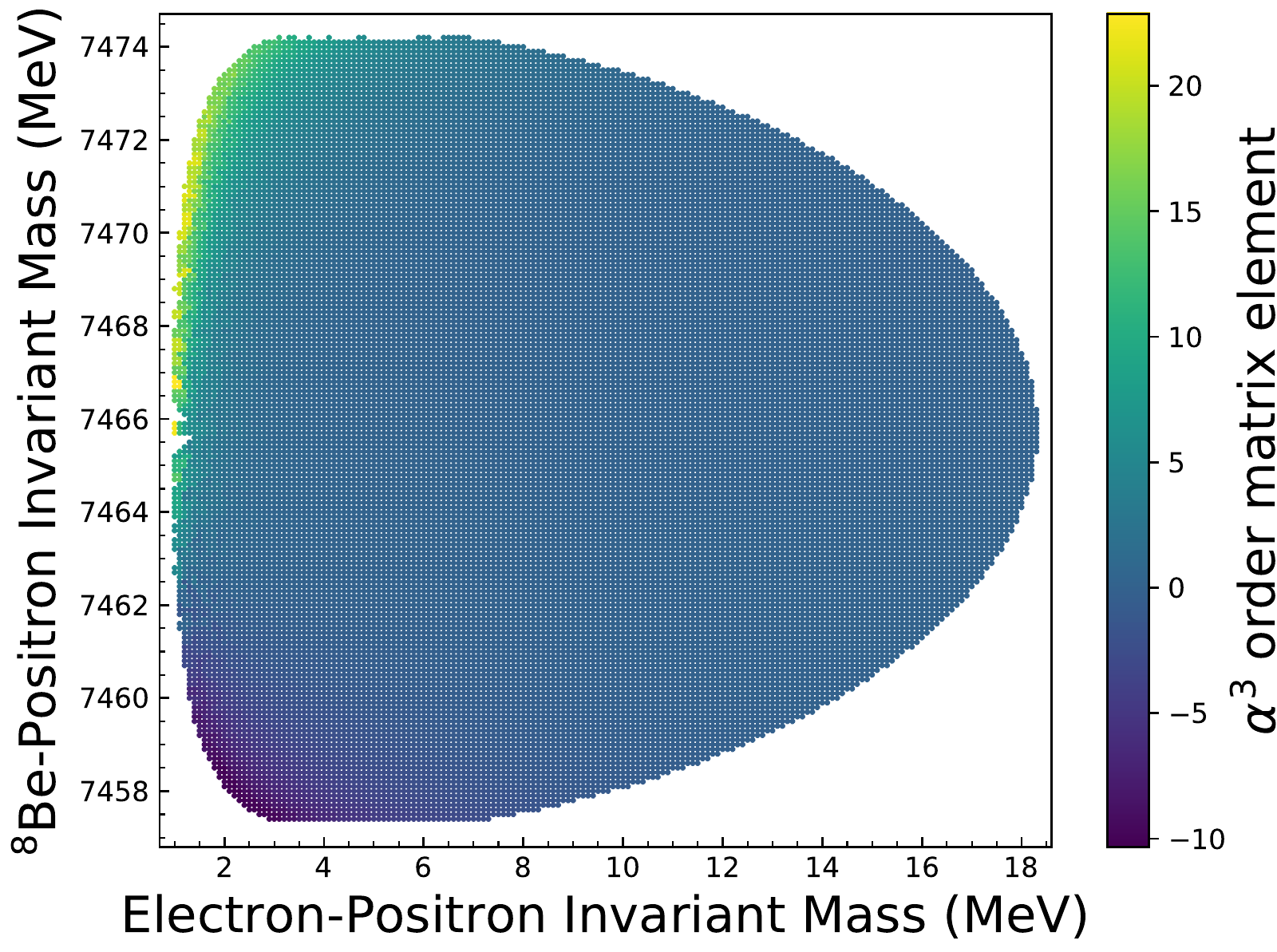} 
       \includegraphics[width=0.48\textwidth]{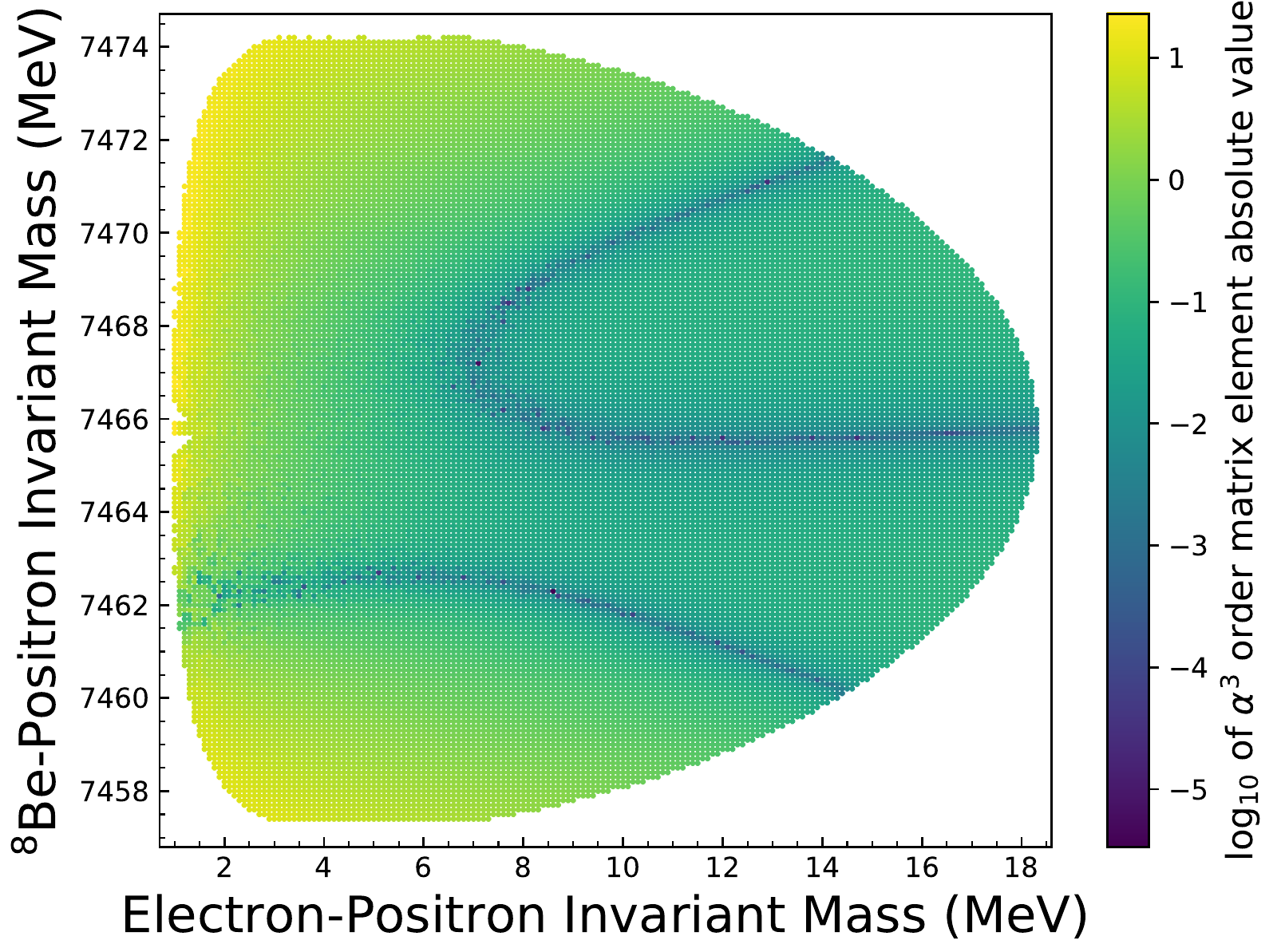} 
     \caption{\label{fig:dalitzLinearMatrix} $\alpha^3$ order matrix element contribution as a function of invariant mass of the $\ee$ system and the $ e^+~\beGround$ system on linear (left) and log (right) scales}
  \label{fig:DalitzME}
  \end{figure}

The phase space of \beGround\ and the electron-positron pair in the final state is shown on Fig.~\ref{figPS1}, demonstrating the characteristic three-body decay behavior. The correlation between the invariant masses in the Dalitz plot plane and the $e^+e^-$ opening angle is shown in Fig.~\ref{fig:angleDalitz}. The matrix elements projected on the 3-body phase space Dalitz plot are shown in Fig.~\ref{fig:DalitzME}.

\end{document}